%% file: ProPro.tex
\documentclass[sigconf]{acmart}
\setcopyright{none}
\renewcommand\footnotetextcopyrightpermission[1]{}
\usepackage{endnotes}
\usepackage{color}
\usepackage{graphicx}
\usepackage{wrapfig}
\usepackage{multirow}
\usepackage{listings}
\usepackage{url}
\usepackage{fixltx2e}
\usepackage{balance}
\usepackage{listings}
\usepackage{verbatim}
\usepackage{subfigure}
\usepackage{stfloats}
\usepackage{amsmath}
\usepackage{tikz}
\usepackage{breqn}
\usepackage[norelsize, linesnumbered, ruled, lined, boxed, commentsnumbered]{algorithm2e}
\usepackage{xcolor}

\AtBeginDocument{%
  \providecommand\BibTeX{{%
    Bib\TeX}}}

\newif\ifdraft
\drafttrue

\def\BibTeX{{\rm B\kern-.05em{\sc i\kern-.025em b}\kern-.08em
    T\kern-.1667em\lower.7ex\hbox{E}\kern-.125emX}}

\definecolor{mGreen}{rgb}{0,0.6,0}
\definecolor{mGray}{rgb}{0.5,0.5,0.5}
\definecolor{mPurple}{rgb}{0.58,0,0.82}
\definecolor{backgroundColour}{rgb}{0.95,0.95,0.92}
\definecolor{amber}{rgb}{1.0, 0.75, 0.0}

\lstdefinestyle{CStyle}{
    backgroundcolor=\color{backgroundColour},   
    commentstyle=\color{mGreen},
    keywordstyle=\color{magenta},
    numberstyle=\tiny\color{mGray},
    stringstyle=\color{mPurple},
    basicstyle=\ttfamily\footnotesize,
    breakatwhitespace=false,         
    breaklines=true,                 
    captionpos=b,                    
    keepspaces=true,                 
    numbers=left,                    
    numbersep=5pt,                  
    showspaces=false,                
    showstringspaces=false,
    showtabs=false,                  
    tabsize=2,
    language=C
}

\newif\ifdraft
\drafttrue
\ifdraft
\newcommand{\zhao}[1]{{\textcolor{cyan}    { ***Zhao:      #1 }}}
\newcommand{\shreyas}[1]{{\textcolor{magenta}    { ***Shreyas:      #1 }}}
\newcommand{\revision}[1]{{\textcolor{blue}    {      #1 }}}
\newcommand{\shivaram}[1]{{\textcolor{red}    { ***Shivaram:      #1 }}}
\newcommand{\pengfei}[1]{ {\textcolor{green}    { ***Pengfei:      #1 }}}
\newcommand{\qiyang}[1]{ {\textcolor{amber}    { ***Qiyang:      #1 }}}
\else
\newcommand{\zhao}[1]{}
\newcommand{\shreyas}[1]{}
\newcommand{\revision}[1]{}
\newcommand{\shivaram}[1]{}
\newcommand{\pengfei}[1]{}
\newcommand{\qiyang}[1]{}
\fi

\newcommand{\up}{\vspace*{-0.5em}}

\newcommand{\name}{Mirage\xspace}
\setcopyright{acmcopyright}
\copyrightyear{2018}
\acmYear{2018}
\acmDOI{XXXXXXX.XXXXXXX}

\acmConference[Conference acronym 'XX]{Make sure to enter the correct
  conference title from your rights confirmation emai}{June 03--05,
  2018}{Woodstock, NY}
\acmPrice{15.00}
\acmISBN{978-1-4503-XXXX-X/18/06}

\begin{document}

\title{Mirage: Towards Low-interruption Services on Batch GPU Clusters with Reinforcement Learning}
\author{Qiyang Ding$^*$}
\affiliation{
    \institution{The University of Texas at Austin} 
    \country{}
}
\author{Pengfei Zheng$^*$}
\affiliation{
    \institution{University of Wisconsin, Madison} 
    \country{}
}
\author{Shreyas Kudari}
\affiliation{
    \institution{The University of Texas at Austin} 
    \country{}
}
\author{Shivaram Venkataraman}
\affiliation{
    \institution{University of Wisconsin, Madison}
    \city{}
    \country{}
}
\author{Zhao Zhang}
\affiliation{
    \institution{Texas Advanced Computing Center} 
    \country{}
}

\begin{abstract}
Accommodating long-running deep learning (DL) training and inference jobs is challenging on GPU clusters that use traditional batch schedulers, such as Slurm. Given fixed wall clock time limits,
DL researchers usually need to run a sequence of batch jobs and experience long interruptions on overloaded machines.
Such interruptions significantly lower the research productivity and QoS for services that are deployed in production. 
To mitigate the issues from interruption, we investigate a set of statistical learning and reinforcement learning (RL) techniques, including random forest, xgboost, Deep Q-Network, and policy gradient to design a proactive provisioner using production job traces from three GPU clusters.
We follow the standard machine learning practice by partitioning each job trace into training and validation subsets, then train each model using the training subset and evaluate the generality using the validation subset. 
We introduce \name{}, a Slurm-compatible resource provisioner that integrates the candidate RL methods.
Our experiments show that the \name{} can reduce the interruption by 17-100\% and safeguard 23\%-76\% of jobs with zero interruption across varying load levels on the three clusters.

\end{abstract}

\maketitle
\pagestyle{plain}
\def\thefootnote{*}\footnotetext{Equal contribution to this work}\def\thefootnote{\arabic{footnote}}
\input{introduction.tex}

\section{Preliminaries}
\label{sec:back}
Deep learning applications have been increasingly popular on GPU clusters.
In this section, we discuss the long-running training and inference service demanded by scientists in supercomputers and present the preliminaries for \textbf{deep Q-learning \cite{mnih2015human}, policy gradient \cite{silver2018general} and mixture of experts \cite{yuksel2012twenty}}.

\subsection{Long Running Deep Learning Applications}
\label{sec:back:trend}
As scientists and DL practitioners investigate more complex problems, they often utilize larger models and datasets.
For example, the 345 million parameter BERT~\cite{devlin2018bert} model takes 5 days to train with 8 NVIDIA A100 GPUs on the English Wikipedia dataset~\cite{pauloski2021kaisa}.
The 175 billion parameter GPT-3 model is estimated to take 34 days to train with 1,024 NVIDIA A100 GPUs~\cite{narayanan2021efficient}.
Scientists are also deploying trained models for real-time data processing or object detection.
Some examples include transient celestial object detection~\cite{ivezic2019lsst}, point-scanning electromagnetic imaging super-resolution~\cite{fang2021deep}, and disease detection in digital agriculture~\cite{mohanty2016using}.
These applications require running long inference jobs as a service, which is problematic since modern GPU clusters frequently impose 48- or 72-hour limits to ensure responsiveness.

\subsection{Deep Q-Learning}
\label{sec:back:rl}

Reinforcement learning (RL) is a machine learning method based on trial and error. At a time step $t$, an RL agent takes an action $A_t$ to react to the current system state $S_t$, and thereby, shifts the system to the next state $S_{t+1}$. The agent receives an immediate reward $R_t$ for triggering the transition, and this process of making decisions and collecting rewards repeats to until a terminal step $T$. The RL agent is trained to learn an optimal that maximizes the accrued rewards within the time horizon ($t=1,\dots, T$). When the learned policy is non-deterministic and the state transition is stochastic, the \underline {trajectory} of produced states, actions and rewards is uncertain and we denote it with sequence of random variables $S_1, A_1, R1$, ...,  $S_T, A_T, R_T$. 

At time step $t$, we use function $Q_{\pi}(S, A)$ to indicate the the expected return (cumulative reward) starting from $t$. $\pi$ itself is function that maps an encountered state to its decided action, and the discounting factor $0\le\gamma\le1$ translates future rewards to present values. 
\begin{align}
    Q_{\pi}(s, a) = E_{\pi}[\sum_{k=1}^{T} \gamma^{k}R_{t+k+1}|S_t=s, A_t=a]
\end{align}
with Monte Carlo simulations, one can rollout different trajectories from a policy, and then, compute an average return for each visited state as a sample estimate of $Q(S, A)$. This would converge if every state and action is visited infinite times. Learning can also happen online by bootstrapping with the $Q$ values of the next state in the sequence,, which is referred to as Q-learning \cite{mnih2015human}:

%
\begin{equation}
\begin{aligned}
    Q_{k+1}(s, a) \leftarrow Q_{k}(s, a) + \alpha(R+ \gamma \max_{a'}Q_k(s', a')-Q_k(s, a))
\end{aligned}
\end{equation}
$s'$ is the next state transitioned from $s$, and $k$ represents the iteration of learning that updates $Q$. The Q-values are captured in a tabular manner but this could get cumbersome for large state-spaces. Recent studies \cite{Silver2016, silver2018general} use a deep neural network with weights $\theta$, namely a deep Q network (DQN),  to represent $Q$ and its loss function and gradients for the $k$-th iteration are defined as below. Rather than computing the full expectation, the loss function can be optimized with mini-batch Stochastic Gradient Descent (SGD).

\begin{align}
    L(\theta_{k+1}) = E[(R+\gamma {max}_{a'}Q(s',a';\theta_{k})-Q(s,a;\theta_{k+1}))^2]
\end{align}
\begin{equation}\label{eqn:dqn_gradient}
    \begin{aligned}
        \nabla_{\theta_{k+1}}L(\theta_{k+1}) = E[R+\gamma  {max}_{a'}Q_k(s',a';\theta_{k})\\-Q(s,a;\theta_{i})\nabla_{\theta_{k}}Q(s,a;\theta_{k+1})]
    \end{aligned}
\end{equation}


\input{pg_formulation}
\input{moe_formulation}

\input{job-analysis}

\input{model-design.tex}

\input{system-design.tex}

\input{experiments.tex}

\section{Related Works}
\label{sec:related}
Facilitating long-running services on batch GPU clusters is relatively new in the area, due to the recently rising adoption of machine learning and deep learning in scientific research.
National computing centers are deploying several large scale GPU dense supercomputers, such as Perlmutter at NERSC, Polaris at ALCF, and Frontier at OLCF. 
Similarly, industry leaders such as Microsoft, Meta, and Tesla deploy their own GPU supercomputers for internal DL-driven research~\cite{tesla, langston2020microsoft, meta2022}.
All these machines have $O(10^4-10^5)$ high end GPUs.
Most of these machines use batch schedulers such as Slurm and Cobalt to manage compute resources.
Several NSF (National Science Foundation) funded AI institutes are devoted to enabling AI/ML/DL in application domains such as agriculture and ecology.
Supporting inference as long-running services is as important as model training, and it is challenging for existing schedulers to support such services with low interruption.


On the other hand, reinforcement learning has made breakthroughs that match or exceed human capability in many areas such as game playing~\cite{silver2018general}, Tokamak reactor control~\cite{degrave2022magnetic}, and protein folding~\cite{jumper2021highly}.
In one way or another, researchers formulate these problems as stochastic control problems then use the classic agent-environment paradigm and evaluate a set of RL techniques of SARSA, deep Q networks, online/offline policy, actor-critic and many more.
In the field of high performance computing and distributed computing, researchers have explored the feasibility of RL in scheduling problems~\cite{zhang2020rlscheduler, orhean2018new, mao2019learning} to maximize certain performance metrics and file system configuration~\cite{li2017capes} to adapt to online I/O workloads.
In this paper, we view the resource provisioning problem as a control problem and train a transformer-based neural network that predicts the expected interruption/overlap, then we use the deep Q network and policy gradient methods to make job submission decisions.

\section{Conclusion and Future Work}
\label{sec:conc}
We examine a suite of ensemble learning and reinforcement learning methods to build \name{}, a proactive resource provisioner towards facilitating long-running DL training and inference services on batch GPU clusters with low interruption/overlap.
\name{} is trained and validated with months-long job traces on three productive GPU clusters.
Our experiment shows that \name{} can enable 23\%-76\% more jobs with zero interruptions, especially when the queue wait time is long.
\name{} effectively reduces the interruption by 17-100\% across the three clusters compared to the reactive baseline.
In future, we will examine the generality of the proposed methods on GPU cluster with a much larger size.

\balance
\bibliographystyle{ACM-Reference-Format}
\bibliography{ProPro}


\end{document}

%% file: introduction.tex
\section{Introduction}

DL practitioners and researchers are increasingly leverage large-scale GPU clusters to train big models.
For example, some scientists are training the 345 million parameter BERT~\cite{devlin2018bert} or the
175 billion parameter GPT-3~\cite{brown2020language} models for various DL tasks and these large
networks can take $O(10^3)$ GPUs for days and even months to train~\cite{zhang2022opt, sharir2020cost}. 
Others deploy ML models to perform inference for classifying celestial objects and detect
Type Ia supernovae on streaming data in real-time~\cite{ivezic2019lsst}. These jobs are long-running as they continuously process incoming data.
As a result, computing centers are experiencing large increases in execution time and wait time, a trend that is expected to continue.  

\begin{figure}[!t]
  \begin{center}
    \includegraphics[width=85mm]{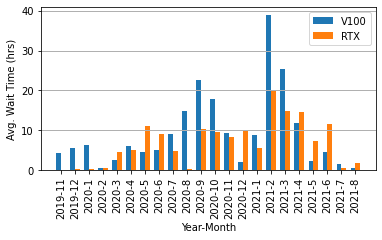}
    \caption{Average Queue Wait Time on the V100 (TACC Longhorn) and RTX (TACC Frontera) GPU Cluster.}  \label{fig:waittime}
  \end{center}
  \vspace{-1ex}
\end{figure}

In an effort to ensure responsiveness and fairness, computing centers enforce a fixed wall clock time limit for jobs running on GPU clusters. 
For example, TACC Longhorn has a 48 hour limit, while the NERSC Perlmutter supercomputer has a 12 hour limit.  
Since the runtime of DL training is much longer than these limits (e.g.,pre-training the 20 billion GPT-NeoX models takes 96 Nvidia A100 GPUs for 30 days~\cite{gpt-neox-20b}), scientists resort to running consecutive jobs. 
That is, just before the time limit expires, they checkpoint model training and then resume training by submitting a second job. 
Specifically, this is usually done in an automatic way by submitting an array of jobs to the scheduler, e.g., SLURM, which does not start accumulating the priority of a dependent job until the previous one completes~\cite{SlurmFactor}. 
Users can experience long wait time by submitting consecutive jobs reactively
(e.g., up to 40 hour wait time on a V100 cluster in February 2021, as shown in Figure~\ref{fig:waittime}). 
Thus, the existing \emph{reactive} approach inevitably introduces interruptions, limits the quality of service (QoS) and hurts responsiveness.


A straightforward way to facilitate long-running services is to dedicate a group of compute nodes and relax the wall clock time, such as the real-time queue on NERSC Perlmutter~\cite{Perlmutter-policy}.
This approach requires policy and scheduler changes system-wise and it is prune to low machine utilization, responsiveness, and unfairness~\cite{hindman2011mesos}, so the real-time queue is via special request.
A second way is to predict the queue wait time, as in classic works~\cite{smith1999using, nurmi2006evaluation, nurmi2007qbets}, then to submit subsequent jobs proactively as advised by the prediction.
However, an empirical study showed that prior approaches
only have an average accuracy of 20-72\%, and at least 91\% of predictions are incorrect~\cite{sonmez2009trace}.
The inaccurate queue wait time prediction attributes to the randomness in job arrival and completion time (a detailed study is in \S\ref{sec:jobs}).
In this work, we take a different approach to reducing the interruption by designing a \emph{proactive} job provisioner that learns the queue state changes using historical job traces via reinforcement learning techniques. 
Researchers have examined RL methods for job scheduling in High-performance Computing~\cite{zhang2020rlscheduler, orhean2018new, mao2019learning} to maximize global performance metrics such as machine-wise utilization or fairness among all users. In contrast, the scope of this work is to minimize the overall time-to-solution for an individual user. 

Specifically, we formulate the proactive resource provisioning problem using the \emph{reinforcement
learning} (RL) paradigm. The proactive resource provisioner functions as the agent. The environment includes the cluster and 
jobs. The machine states and queue states form the environment at a specific time.
The provisioner can extract the machine states and queue states from the scheduler (e.g., Slurm).
The action that the resource provisioner (agent) takes is to either submit or do nothing.
A reward can be determined by the resulting interruption or overlap for a series of actions taken during the execution.

In particular, we investigate the Deep Q-Network (DQN) and policy gradient methods using two network architectures, transformers~\cite{VaswaniShazeer17-attention} and Mixture-of-Experts, as the state-action value function and the policy function, respectively.
We design \name{}, a flexible RL framework with changeable network architecture and RL methods, which interacts with a low overhead Slurm simulator that has been validated to be sufficiently close the production Slurm deployments.

To examine the \underline{effectiveness} and \underline{generality} of the candidate methods, we evaluate \name{} using traces from \emph{three} production GPU clusters. This includes 20-month long traces from TACC Longhorn (a 88-node V100 cluster) and Frontera (a 84-node RTX cluster) and a five-month long trace from Lonestar6 (a 76-node A100 cluster). 
We refer to these three CPU clusters as V100, RTX, and A100. 
Following standard ML practice, we partition the job traces of each cluster into training and validation ranges with ratio of 80:20.
For each cluster, the RL method is trained on the training partition and then validated on the validation range to examine generality.
Our results show that, for HPC workloads, under a medium to high level of cluster load, \name{} safeguards 23\%-72\%, 35\%-72\%,  and 40\%-60\% jobs with zero interruption on the V100, RTX, A100 cluster, respectively.  
We also observe that \name{} significantly outperforms the two heuristic-based strategies in reducing the average interruption by 25-53\%, 21-44\%, 77-100\%, when machines are heavily loaded across the three job traces.

By comparing across all eight methods (see \S\ref{sec:expr}), the MoE+DQN (Mixture-of-Experts with Deep Q-Learning) and transformer+PG (policy gradient agent with transformer) methods outperform others.
When the machine is heavily loaded, transformer+PG has a 16.9-55.9\% lower interruption compared to MoE. 
Its aggressiveness also pays a 1.8-2.6x higher overlap when the machines are lightly loaded.
Given the balanced performance, \name{} uses MoE+DQN as its default model.
We also provide an option for users to use transformer+PG if they work on a heavily loaded machine.

One should not misunderstand the effectiveness scope of the examined RL methods. 
These methods are trained with a cluster job trace, and they are only effective on that cluster.
The generality lies in the methods but not in the models, in the sense that one need to train the models with the particular job trace of the machine that he/she wants to run the proactive provisioner on.

Mirage, the trained models, Slurm simulator, and model training code is available at \url{https://github.com/zhaozhang/Mirage}.
The main contributions of this paper include:
\begin{itemize}
    \item The design and evaluation of DQN and policy gradient methods with transformer and MoE network architecture that significantly reduces interruptions caused by heavy machine load.
    \item The flexible Mirage framework with changeable RL methods and network architecture.
    \item The Slurm simulator that can support job trace sampling and replaying with low overhead.
    \item The generality of the proposed method on three disticnt GPU clusters.
    \item The open source implementation of the simulator and RL-based provisioner.
\end{itemize}

The applicability of Mirage is beyond DL training and inference service. 
Scientists can also use this tool for long-running simulations or data analysis workflows for shorter experiment turnaround.


The rest of the paper is organized as following: We discuss the trend of DL jobs and the basic of RL in \S\ref{sec:back}. 
We analyze the job traces and present the results in \S\ref{sec:jobs}.
The formalization of resource provisioning as DL is discussed in \S\ref{sec:rl} and the implementation details are in \S\ref{sec:impl}.
We present and discuss the experiment design and results in \S\ref{sec:expr}.
A review of RL research and its application in HPC is presented in \S\ref{sec:related}.
Finally, we conclude in \S\ref{sec:conc}.

%% file: pg_formulation.tex
\subsection{Policy Gradient}\label{sec:pgformulation}

Deep Q-Networks (DQN) is a value-based RL method that is known for a few limitations. First, DQN lacks a direct representation of the policy, and policy inference requires solving an optimization problem $\text{argmax}_{a}Q(s,a)$ and this challenges when the action space is high-dimensional. Second, DQN does not automatically trade off exploit and exploration, and usually relies on human-crafted strategies such as the $\epsilon$-greedy strategy \cite{mnih2015human} to guarantee exploration. However, manually tuning the $\epsilon$ threshold is difficult. To overcome these limitations, policy gradient \cite{silver2018general, Silver2016} is widely used; it parameterizes a policy with a deep neural network and directly outputs the probability of the action to take. Moreover, policy gradient can autonomously arbitrate exploit and exploration. 

We briefly formulate policy gradient as below. First, let $\tau$ denote a random trajectory that comprises of a sequence of state, action, reward triples, i.e., $s_1^{\tau}, a_1^{\tau}, r_1^{\tau}, \dots, s_T^{\tau}, a_T^{\tau}, r_T^{\tau}$, to until the terminal time step $T$; let $\pi_{\theta}(\tau)$ denote the probability distribution of $\tau$ when it is realized with a stochastic policy $\pi$. $\pi$ is implemented using a deep neural network with parameters $\theta$; let $r(\tau)$ represent the cumulative reward $\sum_{t=1}^{T}r_t$ and let $J(\theta)$ represent the expectation of $r(\tau)$, i.e., $J(\theta)$=$\mathbb{E}[\sum_{t=1}^{T}r_t^{\tau}]$. Note that, unlike value based methods, policy gradient directly maps a state $s_t$ to its decided action $a_t$ with (softmax) probability $\pi_{\theta}(a_t|s_t)$. The policy gradient theorem states that 
\begin{equation}\label{eqn:pgformulation_1}
    \begin{aligned}
        \nabla_{\theta}J(\theta)=\mathbb{E}_{\tau\sim\pi_{\theta}(\tau)}[\nabla_{\theta}log\pi_{\theta}(\tau)r(\tau)]
    \end{aligned}
\end{equation}
, which can be approximated with Monte Carlo rollout,
\begin{equation}\label{eqn:pg_gradient}
    \begin{aligned}
        \nabla_{\theta}J(\theta)\approx\frac{1}{N}(\sum_{t=1}^{T}\nabla_{\theta}log\pi_{\theta}(a_t^{\tau}|s_t^{\tau})(\sum_{t=1}^{T}r_t^{\tau})
    \end{aligned}
\end{equation}
, and the policy network can be update with gradient ascent and learning rate $\alpha$, i.e., $\theta\leftarrow\theta+\alpha\nabla_{\theta}J(\theta)$, which iteratively converge to a (local) maximum. Note that although policy gradient solves some of the issues of deep Q-learning, it also incur some other issues; its learned policy may converge to a local optimum \cite{pierrot2022diversity}, and in real-world applications, it is found to suffer from high-variances \cite{papini2018stochastic}. As there arguably is no clear winner, we adopt both the DQN and the policy gradient method.

%% file: moe_formulation.tex
\subsection{Mixture of Experts (MoE)}\label{sec:moe_formulation}

Scaling up model capacity is one of the key success of deep learning, as increased model capacity usually renders higher predictive power. There are three learning strategies to enhance model capacity, i.e., scaling up a monolithic model \cite{li2021training}, ensemble learning \cite{Codella2016} and MoE (Mixture of Experts) \cite{yuksel2012twenty}. We try all these three different strategies to build and train RL provisioners, and evaluate them in Section \ref{sec:rl}. Note that we increase model capacity but cautiously prevent over-fitting using cross-validation.

\textit{Scaling up a monolithic model.} One strategy to increase model capacity is to train a huge, monolithic deep neural network that may include hundreds to thousands of layers and billions of parameters \cite{li2021training}. However, training a huge monolithic model is a herculean task. First, training efficacy deteriorates as gradient vanishes when back-propagated through a deep architecture. Residual bypasses and advanced activation functions mitigate this issue but with no complete solution. Second, a huge, complex model has a higher degree of curvature and contains more saddle points and local minima, and gradient descent is more prone to stuck. Third, training a huge neural network is computationally prohibitive. 

\textit{Ensemble learning.} The other strategy to level up model capacity is ensemble learning such as Random Forest (RF) \cite{breiman2001random} and Gradient Boosting Decision Trees (GBDT) \cite{friedman2001greedy}, which average the outputs of multiple learners, each trained on a sub-datasest. However, aggregating a large (infinite) number of weak learners (each underfits the complex data and incurs high biases), increases model complexity but does not effectively increase model capacity. Thus, RF or GBDT usually have difficulties in modeling complex supervised learning tasks such as language modeling. Moreover, classic ensemble learning adopts a static mixture weights for its member learners. However, learners usually perform differently for varied input regions, and the mixture weights should be adaptively different for different input samples.

\textbf{Mixture of Experts (MoE).} We believe the philosophy that instead of training a complex, monolithic model or an ensemble of many weak models, one should train and combine several models of intermediate complexity \cite{hinton_moe}. Following this philosophy, the MoE deep neural networks \cite{yuksel2012twenty} has achieved remarkable success. An MoE neural network contains many experts that share the same network architecture, each of which is trained on a partition of the training samples. There is a mixture (or gating) layer that adaptively routes inputs to their best-fit experts, and the best-fit experts' out are aggregated (averaged) as the final output. This MoE scheme scales up model capacity with no significant increase of computation overhead. We adopt the softmax gating layer \cite{moe_shazeer17}, which computes the weighted average of the Q-values across $E$ different DQN experts.
\begin{equation}\label{eqn:pgformulation_1}
    \begin{aligned}
        Q(s,a)=\sum_{e=1}^{E} G_{\theta}(e)Q_{e}(s,a)\;\;,  G_{\theta}(\cdot)=softmax(x\cdot W_{\theta})
    \end{aligned}
\end{equation}
\color{black}

%% file: job-analysis.tex
\section{Trace Analysis and Data Cleaning}
\label{sec:jobs}

We collected job traces from three production GPU clusters in a national computing center.
V100 has 88 compute nodes, each with four Nvidia V100 GPUs.
RTX has 84 compute nodes, each with four Nvidia RTX 5000 GPUs.
A100 has 76 compute nodes, each with three Nvidia A100 GPUs. 
Specifically, we collect the fields of {\it JobID, JobName, UserID, SubmitTime, StartTime, EndTime, Timelimit, NumNodes}. 
The time span of the job traces are 21, 20, and 4 months, respectively.
These job traces reflect significantly different types of workloads in job arrival time, execution time, node count, node hour consumption, and job queue wait time.
We first analyze the differences quantitatively and then discuss how the job trace data is cleaned before training.

\subsection{Difference in Job Traces}
\label{sec:analysis:diff}
\begin{figure}[h]
  \begin{center}
    \includegraphics[width=85mm]{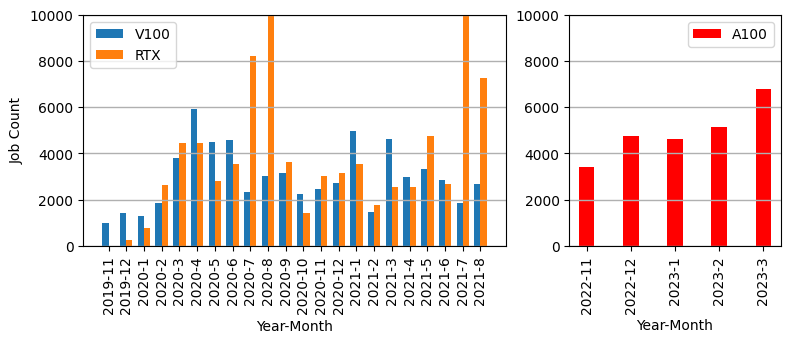}
    \caption{Job Arrival Distribution on the V100, RTX, and A100 Clusters.}
    \label{fig:jobcount}
  \end{center}
  \vspace{-1ex}
\end{figure}

As shown in Table~\ref{tb:job_trace}, the V100, RTX, and A100 traces have 65,017, 175,090, and 17,570 jobs respectively.
Figure~\ref{fig:jobcount} depicts the job count distribution over time.
The average job count is $2,955\pm1,289$, $8,378\pm20177$, and $4,377\pm659$ per month for the V100, RTX, A100 cluster respectively.
As we can see from the figure, there is no clear pattern of job arrival at a month granularity.
It is also worth noting that there are 96,780 short jobs (less than 30 secs) on RTX.
We do not remove such short jobs from our job traces, as they reflect the real machine usage at that time.

\begin{table}[t]
\begin{center}
    \caption{Stats of the Job Traces of V100 and RTX.}
    \begin{tabular}{ | c | c | c | c |}
    \hline
               & V100     & RTX    & A100\\ \hline \hline
    node count    & 88 & 84 & 76\\ \hline
    Start Time & 11/04/2019 & 12/04/2019 & 11/01/2022\\ \hline
    End Time   & 08/20/2021 & 08/19/2021 & 03/31/2023\\ \hline
    Orig. Job Count & 189,899 & 375,095 & 49,997\\ \hline
    Filtered Job Count & 65,017 & 175,090 & 24,779\\ \hline
    \end{tabular}
    \label{tb:job_trace}
\end{center}   
\end{table}

\begin{figure}[h]
\centering
  \subfigure[V100]{
    \includegraphics[width=29mm,trim=1mm 1mm 1mm 1mm,clip]{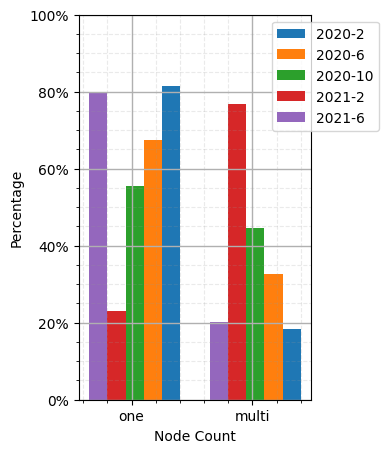}%
    \label{fig:bar-hour-v100}
  }
  \subfigure[RTX]{
    \includegraphics[width=24mm,trim=1mm 1mm 1mm 1mm,clip]{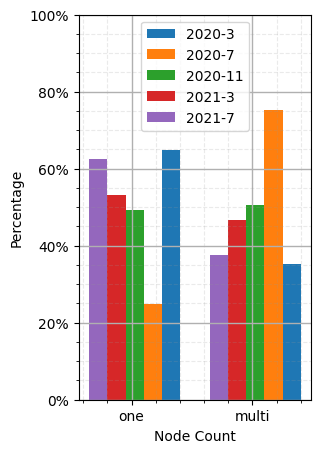}%
    \label{fig:bar-hour-rtx}
  }
  \subfigure[A100]{
    \includegraphics[width=24mm,trim=1mm 1mm 1mm 1mm,clip]{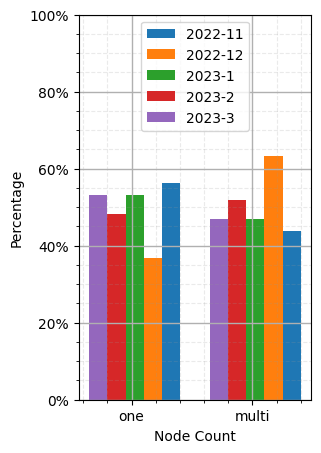}%
    \label{fig:bar-hour-a100}
  }
\vspace{-0.1in}
\caption{Distribution of Node Hour Consumption with Node Count on the V100, RTX, and A100 Cluster.}
\label{fig:node-hour}
\end{figure}

For the node count distribution, 
the average nodes per job is 2.5, 1.3, and 1.6 on V100, RTX, and A100, respectively.
Although multi-node jobs take only a small portion of total job count, their share in node hour consumption is more significant as shown in Figures~\ref{fig:bar-hour-v100}, ~\ref{fig:bar-hour-rtx}, and~\ref{fig:bar-hour-a100}.
For example, in 2021-2 on V100, the percentage of multi-node jobs is 23.4\%, but they take 76.9\% of the total node hours.
Similarly in 2017-11, 12.0\% of jobs are multi-node but take 82.5\% of total node hours.
This observation aligns with the trend of the time-consuming multi-node DL training.

\begin{figure}[h]
\centering
  \subfigure[V100]{
    \includegraphics[width=25mm,trim=1mm 1mm 1mm 1mm,clip]{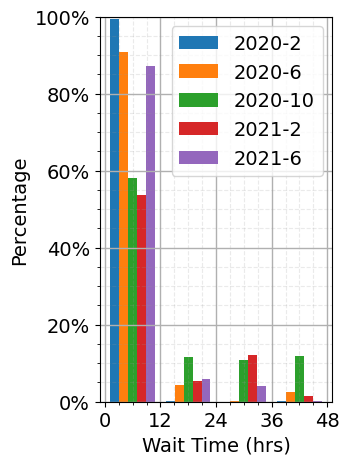}%
    \label{fig:hist-wait-v100}
  }
  \subfigure[RTX]{
    \includegraphics[width=25mm,trim=1mm 1mm 1mm 1mm,clip]{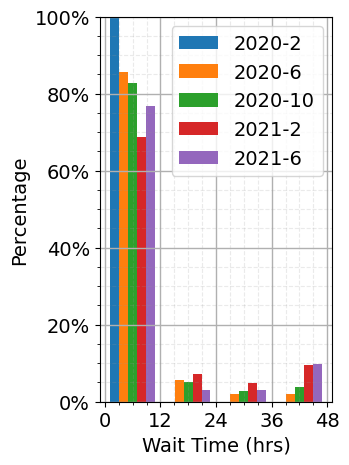}%
    \label{fig:hist-wait-rtx}
  }
  \subfigure[A100]{
    \includegraphics[width=25mm,trim=1mm 1mm 1mm 1mm,clip]{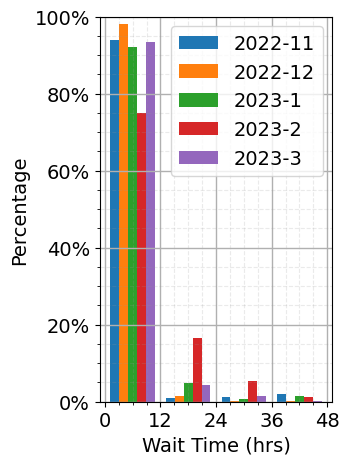}%
    \label{fig:hist-wait-a100}
  }
\vspace{-0.1in}
\caption{Distribution of Queue Wait Time on the V100, RTX, and A100 Cluster.}
\label{fig:wait-time}
\end{figure}

Finally, the statistic most directly relevant to this research, the queue wait time, is depicted in Figure~\ref{fig:wait-time}.
The average queue wait time on the V100 and RTX clusters is shown in Figure~\ref{fig:waittime}.
The distribution from individual months are presented in Figure~\ref{fig:hist-wait-v100}, Figure~\ref{fig:hist-wait-rtx}, and Figure~\ref{fig:hist-wait-a100}.
In 2020-10 and 2021-2, $\sim30-41\%$ of the jobs on V100 have to wait for longer than 24 hours.
On RTX, the percentage of jobs waiting for longer than 24 hours is $\sim12-24\%$.
On A100, 92-98\% of jobs experienced a wait time less than 12 hours across the five-month period except 2023-2, where 26\% of the jobs are waiting for more than 12 hours, and 3\% are waiting for longer than 36 hours.
Long wait time significantly limits the service provided by batch GPU clusters for long running DL training and inference jobs.

\subsection{Data Cleaning}
We manually filter out jobs that 1) request more nodes than the available and 2) sub-jobs that are submitted within one Slurm job.
In the early-production phase of these clusters, all nodes are in the same partition, then they are split into multiple partitions to address the needs of development and test and production run.
Since our job traces are collected in the production partition, there are early jobs that request more nodes than that are available in the partition.
So we remove those jobs from the job traces.
There are also jobs, though recorded in the Slurm database, that are sub-jobs within one Slurm job.
These jobs have an identical prefix followed by the sub-job id.
So we combine them as one single job with the starting time of the first sub-job as the start time, and the end time of the last sub-job as the overall end time.
Table~\ref{tb:job_trace} summarizes the statistics of the job traces.

The machine downtime, e.g., maintenance, leaves blank time ranges in job traces. 
We take the blank time ranges as no users submit any jobs during that time. 
Since the scheduled maintenance is one day every one or two months, it does not impact model training significantly.
Slurm allows users to submit jobs with dependencies, and it releases the next job to the queue upon the finish of the previous job. 
This dependency is not reflected in the job traces, so we take jobs with dependencies as independent jobs submitted at different time.
This approach does not change the queue wait time of the subsequent jobs.

%% file: model-design.tex
\section{Reinforcement Learning for Proactive Provisioning}
\label{sec:rl}

Following the advances in deep supervised and unsupervised learning, reinforcement learning (RL) has embraced deep architectures and has achieved success in solving complicated, human-level control problems such as playing Atari and Go. These recent innovations motivate us to design a smart resource provisioner using RL techniques, and particularly in our framework, deep Q network~\cite{osband2016deep} (DQN) and policy gradient (PG) ~\cite{silver2018general}, which we show can help HPC practitioners perform adaptive, proactive resource provisioning.

We discuss the advantages and disadvantages of DQN and PG in Section \ref{sec:pgformulation}, and implement an RL framework with both DQN and PG policies (cf., Figure~\ref{fig:arch}). Similar to the policy framework in \cite{clipped_proximal_policy_optimization}, instead of building independent value and policy networks, we build the framework with a dual-head architecture, i.e., the V-head or the Q-value head and the P-head or policy gradient head. The two heads share the same foundation model (a transformer or an MoE-transformer model) while have different embedding and output layers. For the Q-head, the embedding layer encodes state-action pairs and the output layer maps their Q-values. For the P-head, the embedding layer encodes only states and the output layer maps to a the probability of different actions, followed by a sampling layer that accordingly samples actions.

To distill high-level features from the cluster state variables and the state transitions, we propose using transformer, which is validated to be effective for a wide-spectrum of ML application including both 
language modeling and generation (GPT-3 transformer \cite{gpt3-github}), and computer vision (ViT, i.e., vision transformer \cite{han2022survey}).
Particularly, we leverage transformer's multi-head attention mechanism to model long-range dependence, predict how cluster states (e.g., queue length and cluster business) transition over time, and thereby, manage interruption (or overlap) proactively.

Note that in addition to RL-based implementations, we also build Mirage with classical statistical models including random forest and gradient boosting decision tree, though results in the experiment section declares a significant gain for the RL models over the statistical model.

\subsection{Encoding Queue, Server and Job States.} 
We represent the system state at each instant $t$ with an $m$-dimensional vector $v_t$. By default, $m$ is 40 and the vector comprises the following variables to encode queue, server and job information. 

\textit{\textbf{(a) Queue State:}} First, queue state includes the number of queued jobs currently waiting for scheduling ($var1$). Second, it includes summary statistics (i.e., 0th, 25th, 50th, 75th, and 100th percentiles) of the sizes (number of requested nodes) of the queued jobs ($var2-var6$). Third, queue state includes summary statistics of the ages (time since submission) of the queued jobs ($var7-var11$). Fourth, it includes summary statistics of the runtime limit (maximum duration that each job is allowed to run in the cluster) of the queued jobs ($var12-var16$). 

\textit{\textbf{(b) Server State:}} First, server state includes the number of actively running jobs across all servers in the cluster ($var17$). Second, it includes summary statistics of the sizes ($var18-var24$), elapsed runtime ($var25-var29)$ and runtime limit ($var30-var34$) of all running jobs.

The model only considers submission time prediction for a single successor job, after the submission of its predecessor job. For example, suppose a user job $J$ is partitioned as four 48 hour long sub-jobs $J1$, $J2$, $J3$ and $J4$. The model only considers optimizing the submission time of the successor sub-job $J2$ after the predecessor sub-job $J1$ is submitted. When $J2$ is submitted per model's decision, $J2$ becomes the predecessor and $J3$ becomes the successor, and so on, until $J4$ is submitted. Therefore, the model maintains a current Predecessor-Successor pair for each group of chained sub-jobs, and at each instant, the state representation also includes the status quo.

\textit{\textbf{(c) Predecessor job state}} and \textit{\textbf{(d) Successor job information:}} The predecessor states includes the size ($var35$), time limit ($var36$), queue time ($var37$), and elapsed runtime ($var38$). The successor information includes its size ($var39$), and time limit ($var40$). The successor job has not entered the cluster yet and thus only static information is included. 
It should be noted that we consider jobs' internal state, such as say an ML job's epoch progress to be private to the user. That is, the agent does not know or require users to annotate or expose any job-specific internal information for training.


\subsection{Encoding Workload and Cluster History} 
\textbf{The state matrix.} The 40-dimensional vector $v_t$ is an abstraction of the queue, cluster and job states at an instant $t$, while $k$ consecutive vectors $v_t$, $v_{t-1}$, $\dots$, $v_{t-k}$ encode both workload (job arrival and completion) and cluster (the queue and server) history of length $k$. Existing RL work, such as playing Altari games \cite{mnih2015human}, adopts a similar approach in constructing model input, within each the last four video frames (last four snapshots of video game) are taken as model input. In parallel, our model input is a $k\times m$-dimensional state matrix $S_t$ at each instant $t$. 
By default, we set $k$ as 144, and periodically record $v_t$ at a default interval of 10 minutes which corresponds to the workload and cluster history of the last 24 hours back-traced from $t$.

\subsection{The State Space and the Action Space} \label{sec:state_action_space}
\textbf{Unifying the state and action space for DQN and PG network.} The action space for our RL provisioner consists of only two different actions: submit (submission) and no-submit (no-submission). The DQN network takes as input both the state matrix and the action to query, while the PG network takes as input only the state matrix. To render a unified model input, we add an ordinal variable that represents the action to query. For the ordinal action variable, 1 represents submission, -1 represents no-submission, for which the DQN network realizes it as an essential input, while the action is only a placeholder and is always 0 when inputted to the PG network. When training the foundation transformer model, we flatten the state matrix to a long vector and concatenate with the additional ordinal action variable. The flattened states (by default) contains 5761 (144$\times$40+1) variables (cf., Figure~\ref{fig:arch_transformer}).

\subsection{Policy Serving} 
\begin{figure}[h!]
\centering
\includegraphics[width=0.8\columnwidth,trim=0mm 0mm 0mm 0mm,clip]{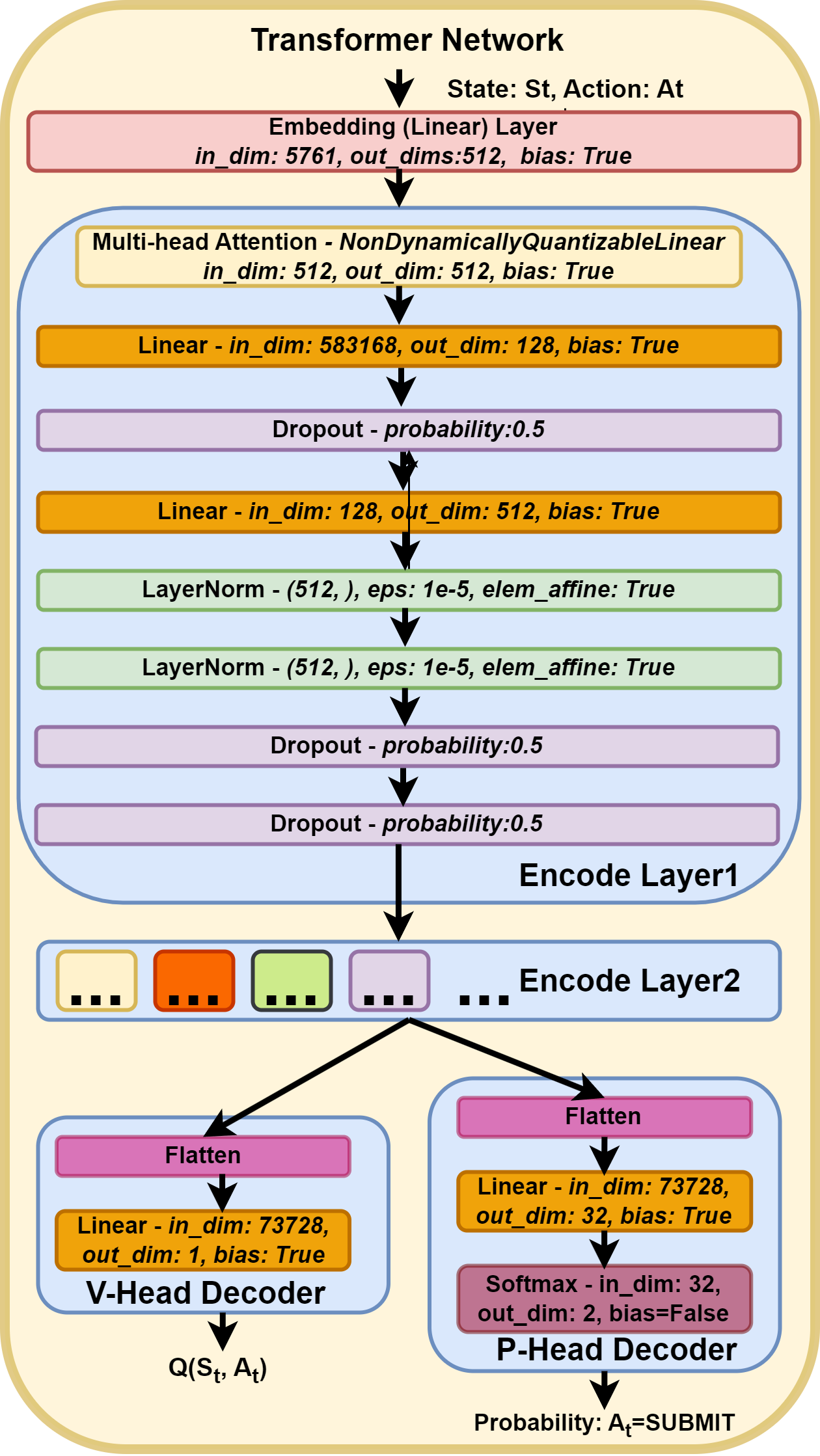}%
\vspace{-0.1in}
\caption{Dual-head Network Architecture - Transformer}
\label{fig:arch_transformer}
\end{figure}

\textbf{Deterministic Policy.} For a learned DQN policy, the neural network is a value function $Q(S_t, a_t)$ that predicts, at the current instant $t$, the expected future gains of submitting versus not submitting the successor sub-job, given the current state and the history encoded in the state matrix $S_t$. For each targeted predecessor-successor pair, after the predecessor is submitted, the DQN model is periodically invoked at an interval of 10 minutes, and will submit the successor sub-job only when the Q-value of submission is larger than that of no-submission. {Non-deterministic policy.} For a learned PG policy, the network outputs directly the probability of submission versus no-submission, and the action to take is randomly sampled from this output binomial distribution (cf., \ref{fig:arch_transformer}).

\subsection{Shaping the Reward}\label{sec:reward_shaping} 
Different HPC practitioners can have their own views of the penalty of overlap and (or) interruption between the predecessor and successor sub-jobs. Performance-sensitive users may consider interruption to have a much larger penalty than overlap, while resource-waste-aversion users may consider overlap to have a larger penalty than interruption. We set two user-configurable penalty coefficients, $e_I$ and $e_O$ , for users to configure their penalty for interruption and overlap, respectively. Given this, we first set the time horizon of actions. Suppose the current instant is $t'$, at which the predecessor sub-job is submitted, and after that, the agent's trained DQN or PG policy produces a sequence of actions $a_{t'+1}$, $\dots$, $a_{t'+T}$. We can see that $a_t=0$ for all $t'+1\leq t<t'+T$, and $a_{t'+T}=1$ as the terminal action for the successor sub-job is submission, while any other actions prior to $a_{t+T}$ are all non-submission actions.

After the successor job is submitted at instant $t'+T$, the agent closely monitors the state of the successor job and when it is dequeued for running at some future instant $t'+T'$, the outcome of how much the overlap or interrupt is revealed. We use $r_I$ to denote an outcome interruption and $r_O$ to denote an outcome overlap.

With the settings above, we shape the reward $R_t$ for actions $a_t$ as below. Note that we use negative penalty to equally represent reward, which means a reward of zero is the best possible reward, and the larger the interruption or overlap penalty is, the smaller the reward.

\begin{equation}
R_t = 
\left\{
    \begin{array}{lr}
        -e_i\cdot r_i, \text{if  INTERRUPT}\\
        -e_o\cdot r_o, \text{if  OVERLAP}
    \end{array}
\right\}, t'+1\leq t\leq t'+T
\end{equation}

If the observed interruption is small, the previous actions are rewarded to
making the right decisions. That is, the previous no-submission decisions are
credited for not causing overlap, and the final submission decision is credited for being proactive and timely. A similar argument applies when the observed overlap is small.

\begin{figure}[h!]
\centering
\includegraphics[width=0.75\columnwidth,trim=0mm 0mm 0mm 0mm,clip]{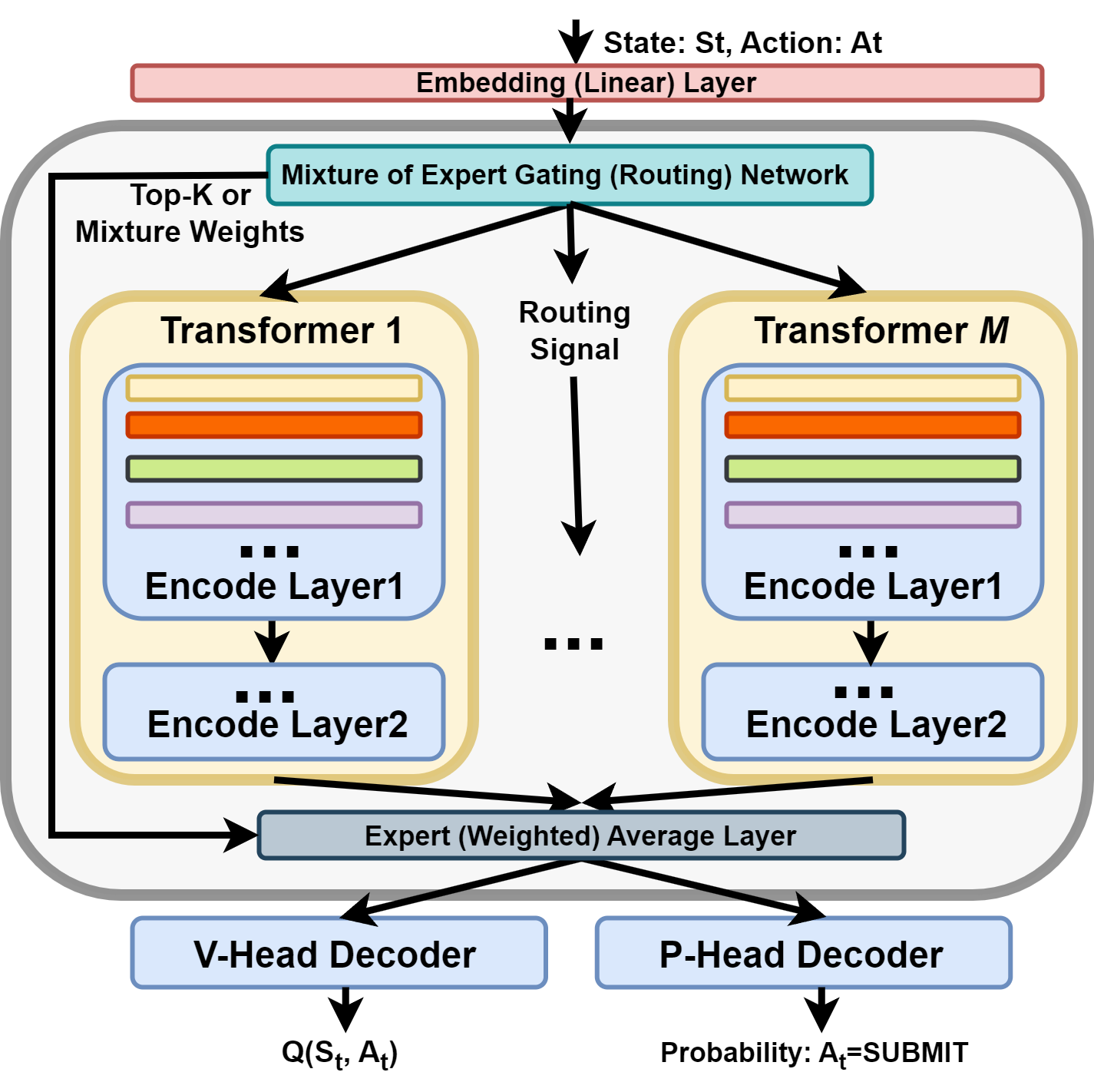}%
\vspace{-0.1in}
\caption{Architecture with the MoE Foundation Model}
\label{fig:arch_moe}
\end{figure}

\subsection{Network Architecture} 
\textbf{The foundation model - transformer.} We build the foundation model with a transformer (Figure~\ref{fig:arch_transformer})) model. Transformers are the state-of-the-art method for sequence modeling and prediction. For example, in NLP tasks like language modeling, a transformer model takes as input a sequence of previous words to predict the sequence of words that follow. This resembles our proactive scheduling task where we parse a history of system (e.g., queue and node) snapshots and forecast the change in system states. The difference lies in that, for language models, the transformer decoder deciphers the intermediate representation into predicted words, while for proactive scheduling, the decoder translates the intermediate representation of future system states into queuing delay, which can be used to compute the resulting interruption (or overlap). Furthermore, system state has both long-range dependencies (e.g., periodic, weekly jobs) and short-range dependencies (e.g., bursty arrival of jobs) over the time horizon. Thus, not all system snapshots in a history window provide dependent, relevant features to forecast the status quo and future. The multi-head self-attention module filters out irrelevant snapshots in history and identifies ones that contribute to prediction. Self-attention is the key technique in Transformer or BERT to improve upon traditional sequence modeling and achieve state-of-the-art results for classification tasks, including language modeling, machine translation and sentiment classification. Thus we propose using self-attention as the key enabler to accurately predict future system states and reinforcement learning rewards.

\textbf{The V-head and the P-head output layers.}  The deep foundation model learns intermediate representation of the policy internal values and the it is either followed by a single linear map to reduce to a scalar output (i.e., the Q-value) of the V-head, or followed by a linear decision layer plus a softmax layer, which output the distribution of action sampling for the P-head (cf., Figure~\ref{fig:arch_transformer}).

\textbf{Hyperparmeter tuning.} The hyperparameters that define the model structure, such as the number of multi-attention heads per transformer encoder layer, are tuned with RayTune~\cite{liaw2018tune}, an auto-ML engine for hyper-parameter tuning. Figure~\ref{fig:arch_transformer} shows the tuned hyperparameters for all network layers. This set of tuned hyperparameters is repeatedly used as the default throughout this study.

\subsection{The MoE foundation model}
As discussed in Section \ref{sec:moe_formulation}, we further augment the capacity of the transformer foundation model with the MoE scheme. Figure~\ref{fig:arch_moe} shows the MoE architecture and the workflow to combine experts. We implement and evaluate both the Top-1 sparse MoE \cite{moe_shazeer17} and the weighted average dense MoE \cite{moe_shazeer17} and found that the Top-1 scheme exhibit inferior provisioning performance in comparison to the weighted average MoE, though it enjoys less computational overhead due to sparse activation. Throughout this study, we omit the results of the Top-1 scheme for brevity.

The intuition behind MoE is similar to that of piecewise polynomial fitting such that individual polynomials focus to fit a small region (partition) of the input space (samples), and all the polynomials integrate to fit a complex function over the entire input space (complete set of samples). As cluster workloads and machine busyness change from weeks to weeks and months to months, we temporally split the cluster log records (training samples) to $M$ (10 by default) fractions to train different expert transformers. Different experts focus on optimizing provision policy for different load levels and different workload mixes.

\subsection{Experience Replay} 
Online RL usually fails for complicated control tasks if there exists correlation between consecutive training samples, which explodes the variance of gradient updates and distorts the a policy's value estimates. We also employ experience replay~\cite{adam2011experience} to break data correlations. Experience instances, each of which includes the submission or no-submission action taken by the trained policy, as well as its resulting reward (interruption or overlap), are stored in a memory pool. The memory pool is random as the experience instances are collected from different time points for each simulation trial, and are randomly shuffled across different simulation trials. Experience instances are then grouped into random mini-batches to perform training of the value (or policy) networks. 

\subsection{Training DQN and PG} The training procedure consists of two phases. For the first phase, we train the foundation model offline while for the second phase, we train the V-head and the P-head decision layers online by running Slurm simulation experiments.

\subsubsection{Offline training}\textbf{a) Sample collection.} We first collects a series of training samples; each episode instantiates an (simulated) experimentation episode that first submits a predecessor sub-job to the cluster at the current instant, and then, submit the successor sub-job at 7 averagely split sample points in a time range between start time (2 days warm-up after the simulation start time) and the end time of predecessor job. The episode lasts until the successor’s delayed reward (i.e., interruption or overlap) can be observed. A single sample, i.e., a (state, action, reward) triple, is collected at the end of the episode and stored into the experience memory pools. When a job is partitioned into $k$ sub-jobs, it can maximally generate $k-1$ training samples (predecessor-successor pairs) for training.
\textbf{b) Foundation model training.} In the offline phase, we pre-train the transformer or MoE-transformer foundation model with supervised learning; using training samples randomly sampled from the memory pool. Each sample includes a state and an observed reward. The input to the foundation model is the flattened state (cf., 
 Section~\ref{sec:state_action_space} and Figure~\ref{fig:arch_transformer}), while the output is the observed reward. We use the Adam optimizer \cite{bock2019proof}.

\subsubsection{Online training} \textbf{a) Online DQN training.}: Online DQN training executes on-policy RL training \cite{mnih2015human} using the gradient formula in Equation~\ref{eqn:dqn_gradient}. Training samples are collected with actions decided by the learned Q-value function. The DQN policy may never submit the successor and this leads to an infinite episode. To prevent such an extreme case, we add a small probability $\epsilon>0$, with which, the DQN leaner randomly chooses actions regardless of the Q-values (i.e., $\epsilon$-greedy strategy \cite{mnih2015human}). \textbf{a) Online PG training.}: Online PG training executes on-policy RL trainig using the gradient formulate in Equation \ref{eqn:pg_gradient}. Note that though the V-head and P-head share the same foundation model, these two heads are trained independently.


%% file: system-design.tex
\begin{figure}[h]
  \begin{center}
    \includegraphics[width=0.85\columnwidth]{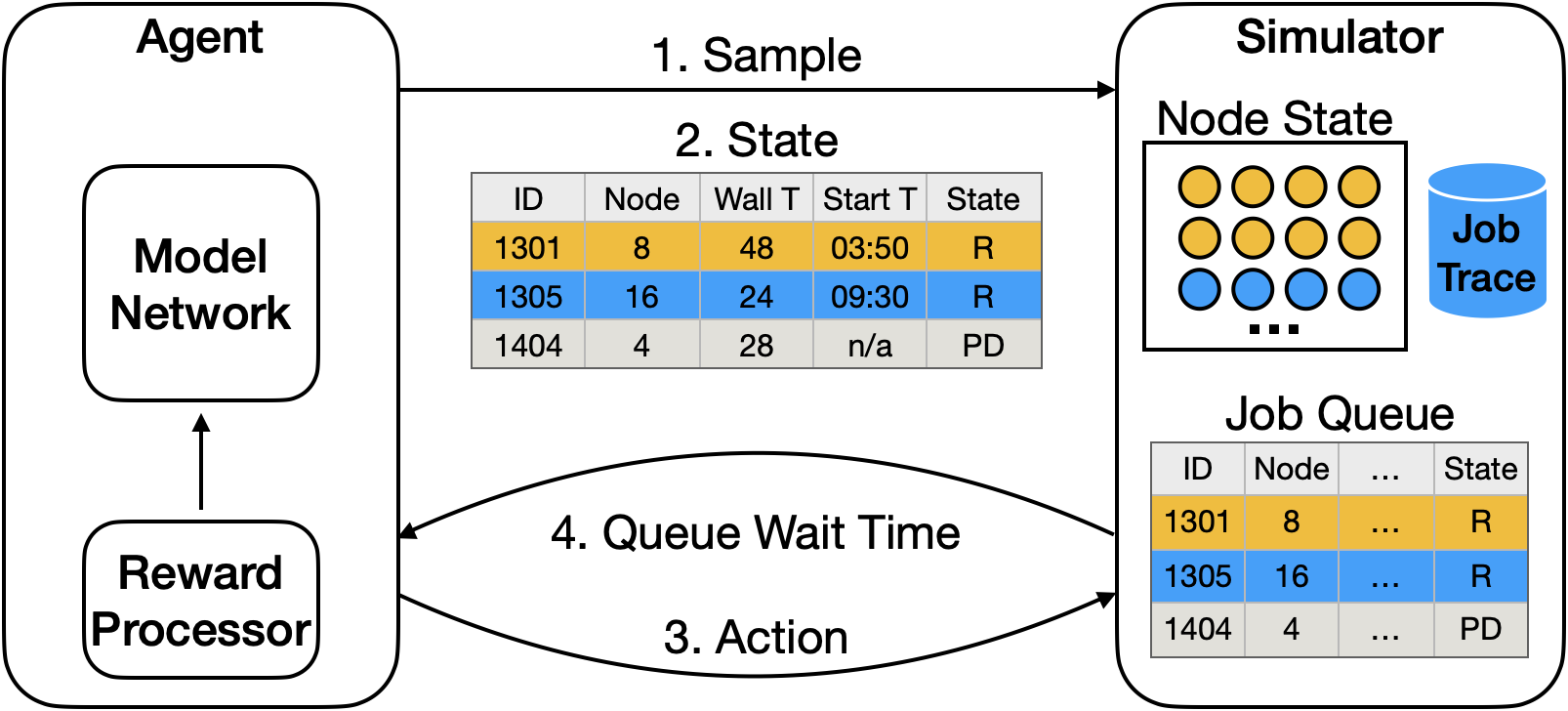}
    \caption{Architecture Overview.}
    \label{fig:arch}
  \end{center}
  \vspace{-1ex}
\end{figure}

\section{System Design and Implementation}
\label{sec:impl}

In this section, we discuss the overall \name{} design and its implementation details.

\subsection{System Architecture}
The overall system architecture of \name{} consists of two major components: the
agent and the simulator, as shown in Figure~\ref{fig:arch}.  It is designed to be generic for both
model-free and model-based learning and handles value function approximation and model training.

At a high level, the agent trains the proactive provisioner and the simulator simulates the Slurm
scheduler using the provided job traces.
The simulator implements the priority-based scheduling algorithm with back-filling~\cite{yoo2003slurm}. 
and mimics the behavior of the production Slurm deployments used in our GPU clusters. 
The simulator loads the jobs trace that we collected over a 20-month period and exposes an interface of {\it sample(), step(), and submit()}.

For each episode, the agent and the simulator will be initialized and run up to the time when the first job should be submitted. 
Then the agent submits the first job by calling {\it submit()}, calls {\it sample()} to acquire the cluster state. 
After that, the agent calls {\it step()} to instruct the simulator to move forward for a period of time, then collects the cluster state again. With sufficient states, the agent evaluates the rewards of submitting the second job, and makes a decision. 
Once a decision has been made, the agent calls {\it submit()} to submit the second job.
Then the simulator runs till the second job starts and returns its final queue wait time that can be used to compute the reward.

The agent has a reward processor that receives the queue wait time from the simulator and converts
it to the reward as required by the RL algorithm. 



\subsection{The Slurm Simulator}
In order to train the agent, and validate the results, we develop a low-overhead Slurm simulator that supports Slurm's core scheduling logic, i.e., backfilling and priority scheduling. There are two main roles fulfilled by our simulator. One is to simulate the workload to create a virtual cluster environment, and the other is to provide support for offline learning and online learning of agents. Since the time taken for each iteration of learning depends on the simulation speed, it is crucial that we have a low-overhead, yet high-fidelity simulator design. Our current build simulates a one month workload within one minute.

Our simulator includes three components, a cluster abstraction, a scheduler abstraction, and a
simulator module that interacts with the agent. The cluster abstraction is used to consume the
workload, while the scheduler determines which jobs will be submitted to the cluster based on its
policy. The simulator module exposes a rich API for running workloads and allows controlling a
number of factors such as running the simulator a specific length of time or running until all the
jobs have been finished etc. The ability for users to customize such factors makes our simulator design general enough to support both offline and online learning.

We evaluate simulation fidelity using an identical workload trace, and compare our simulator with the standard Slurm simulator \cite{slurmsim2022, slurmSimulator1}, which implements the exact Slurm scheduling logic that is used in real-word HPC clusters. We use 5 randomly sampled weeks as input and find that the difference in makespan across the five runs is less than 2.5\%. We also compute the differences in job completion time (JCT) and find that the geometric mean of the difference is no more than 15\% across all runs. Further, our simulator has a 3-26$\times$ lower overhead than the standard Slurm simulator.

%% file: experiments.tex
\section{Experiments and Results}
\label{sec:expr}
To validate the effectiveness and generality of \name{}, we conduct experiments using the job traces of the V100, RTX, and A100 GPU clusters.
We partition each trace in 80:20 ratio for training and validation. 
Specifically, the training split is 11/2019-02/2021 and the validation split is 03/2021-07/2021 on both V100 and RTX. 
For the A100 GPU cluster, the training split is 11/2022-02/2023 and the validation split is 03/2023.

Our evaluation considers the following scenarios:
\begin{itemize}
    \item For each trace, training \name{} with the training data for single node jobs, then using on the validation subset to examine the generality of the proposed method. 
    \item For each trace, training and validating \name{} with 8-node jobs to study the effectiveness of our proposed method on multi-node jobs.
    \item Comparing the RL-based techniques with the statistical methods and the heuristic baselines of reactive provisioning.
\end{itemize}

We implement \name{} using PyTorch 1.11~\cite{NEURIPS2019_9015} and use Ray~\cite{moritz2018ray} to enable distributed data preprocessing. The training is executed on the Lonestar6 supercomputer at Texas Advanced Computing Center.

We use two heuristic baselines: first is {\bf reactive}, which submits the subsequent job upon the completion of the current one.
\textbf{\textit {It is worth noting that, the reactive baseline is what researchers usually use as a common practice~\cite{SlurmFactor}.}}
The second baseline is {\bf avg}, which is derived by monitoring the average queue wait time $T_{avg}$ and submitting the second job $T_{avg}$ time units before the first job finishes. 
The second group of methods are ensemble methods, i.e. {\bf XGBoost}~\cite{chen2016xgboost} and {\bf Random Forest}~\cite{breiman2001random}.
XGBoost combines multiple decision trees and leverages gradient boosting to make predictions.
Random Forest also combines multiple decision trees to output a single output via averaging or voting.
The third group of methods are RL-based methods. We treat transformer and MoE as foundation models, and examine both the DQN and policy gradient learning methods. 
So there are four combinations in total: {\bf \{transformer, MoE\} $\times$ \{DQN, PG\}}. 

For all experiments, we present the evaluation results under different cluster loads. We characterize cluster busyness with the reactive queue wait time measured in the baseline and define three categories: longer than 12 hours (high load), between two and 12 hours (medium load), and within two hours (light load).

\subsection{Single-node Evaluation}
\label{sec:exp:single-val}

\begin{figure*}[]
  \subfigcapmargin=0.1in
  \subfigure[Heavy Load]{
    \captionsetup{width=.1\linewidth}
    \includegraphics[width=3.25in,clip]{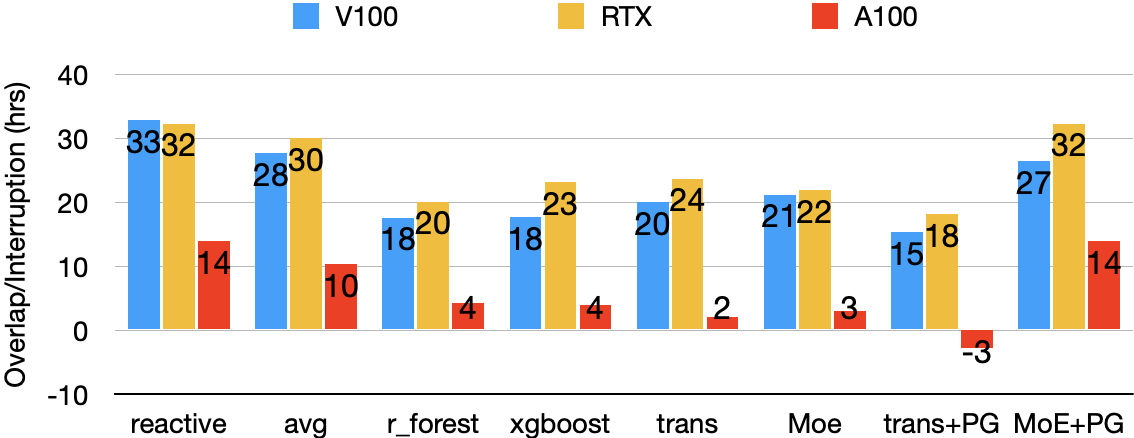}%
    \label{fig:one-heavy}
  }
  \subfigure[Medium Load]{
    \includegraphics[width=3.25in,clip]{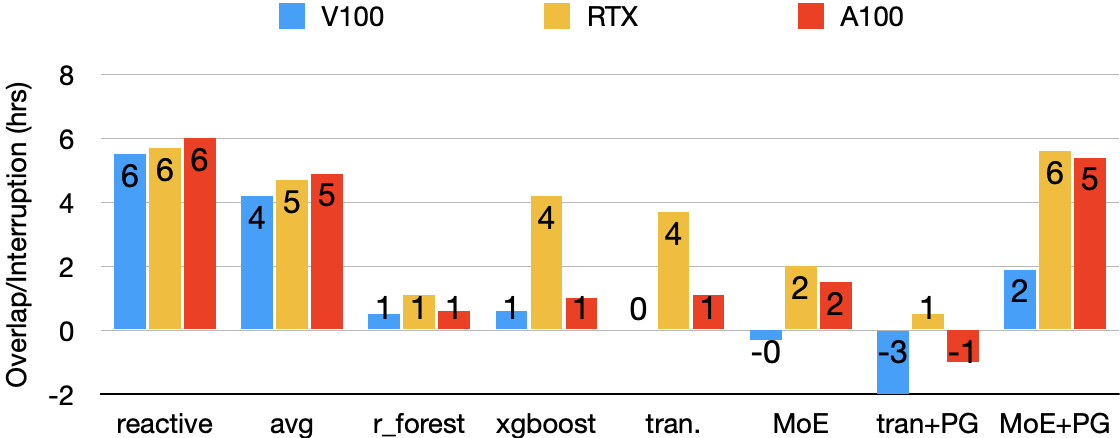}%
    \label{fig:one-medium}
  }
  \caption{Average Interruption of a Pair of 48-hour Single-node Jobs on the V100, RTX, and  A100 GPU Clusters.}
  \label{fig:one-node}\up\up
\end{figure*}

In this experiment, we train each model by uniformly sampling the training time range and ran a pair of consecutive single-node jobs for 48 hours. 
Then we use the same sampling method in the validation range to examine generality.
Single-node jobs are commonly for DL inference service. 
Figure \ref{fig:one-node} shows the average interruption/overlap of all techniques on the three GPU clusters.

As shown in Figure \ref{fig:one-heavy}, when the cluster is heavily loaded, i.e., the reactive queue time exceeding 12 hours, random forest, XGboost, transformer, MoE, MoE+DQN and transformer+PG all show improved interruption with an average reduction of 44.1\%, 33.7\%, and 84.7\% on the V100, RTX, and A100 cluster compared to the reactive baseline, respectively.
MoE+PG is not as effective as other ensemble learning and RL methods. 
By comparing the training loss of MoE+PG with transformer+PG, MoE+PG overfits with equally low training loss with transformer+PG but a much lower validation performance.

We see a similar pattern when the clusters are with medium load in Figure~\ref{fig:one-medium}. 
The ensemble learning techniques, DQN techniques, and transformer+PG show sigfinicant interruption reduction. 
Across the three clusters, Transformer+PG has the lowest interruption followed by MoE.

\subsection{Multi-node Evaluation}
\label{sec:exp:multi-val}
\begin{figure*}[]
  \subfigcapmargin=0.1in
  \subfigure[Heavy Load]{
    \captionsetup{width=.1\linewidth}
    \includegraphics[width=3.25in,clip]{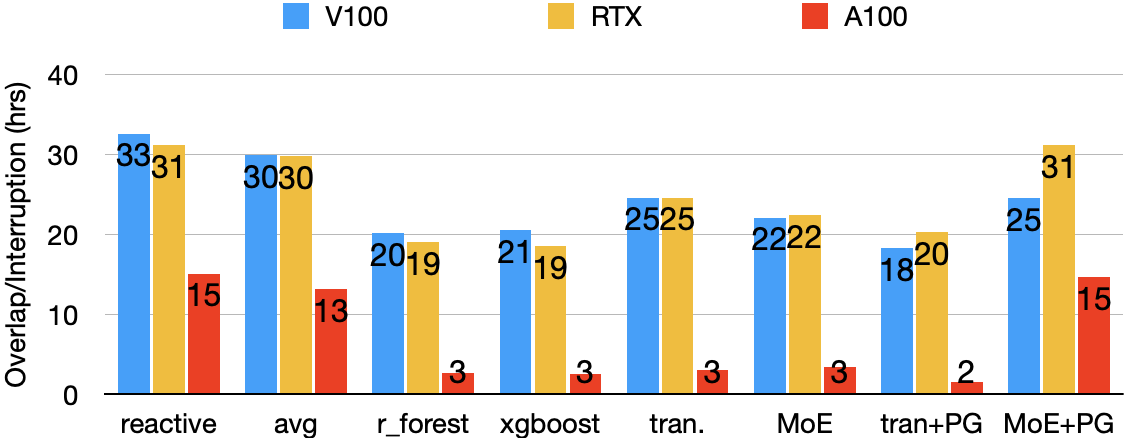}%
    \label{fig:eight-heavy}
  }
  \subfigure[Medium Load]{
    \includegraphics[width=3.25in,clip]{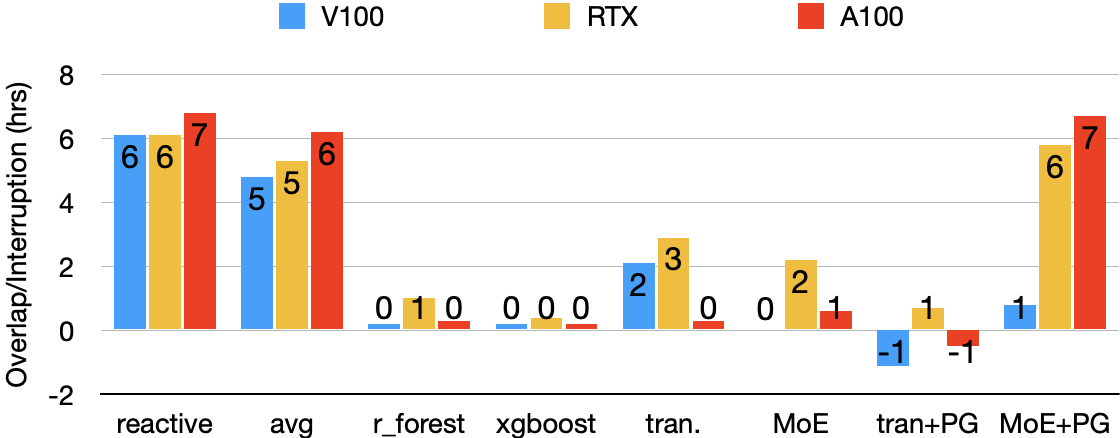}%
    \label{fig:eight-medium}
  }
  \caption{Average Interruption of a Pair of 48-hour Eight-node Jobs on the V100, RTX, and A100 GPU Clusters.}
  \label{fig:eight-node}\up\up
\end{figure*}

In this experiment, we examine the effectiveness of candidate methods using a pair of eight-node jobs, which is a representative scale across the three clusters.
Figure~\ref{fig:eight-node} illustrates comparison with heavy and medium loads.

With heavy load as shown in Figure~\ref{fig:eight-heavy}, the XGBoost and Random Forest methods show promising reduction of interruption by 37.5\%, 40.0\%, and 82.5\% across the three clusters.
MoE+DQN reduces the interruption by 32.2\%, 28.2\%, and 77.5\%, which is slightly behind the ensemble learning methods.
Transformer+PG has an improvement by 43.9\%, 34.9\%, and 90.1\%, which shows the best results if we compare with an average.

With medium load as shown in Figure~\ref{fig:eight-medium},
We observe that ensemble learning methods almost completely eliminate the interruption.
Transformer+PG shows a similar result.
All these interruption elimination is at a cost of overlap when the machines are lightly loaded.
We will discuss this in next section.

Now, comparing the effectiveness of MoE+DQN and transformer+PG methods across clusters, we see that they are more effective on the A100 cluster than on the V100 and RTX clusters.
This is because the A100 job trace has less noisy jobs (those request hours of wall clock time but run for only 30 seconds), as we discussed in \S\ref{sec:analysis:diff}.
The experiments in \S\ref{sec:exp:single-val} and \S\ref{sec:exp:multi-val} with the A100 cluster clearly show the effectiveness of the RL-based methods, especially with a clean job trace.

\label{sec:exp:size-val}

\subsection{Overlap with Low Machine Load}
\label{sec:exp:overlap}
\begin{figure*}[]
  \subfigcapmargin=0.1in
  \subfigure[One Node]{
    \captionsetup{width=.1\linewidth}
    \includegraphics[width=3.25in,clip]{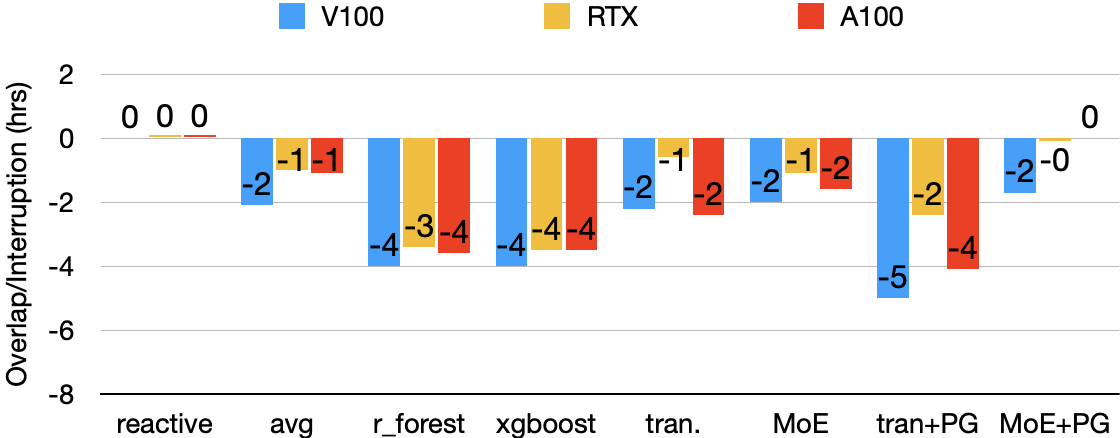}%
    \label{fig:one-low}
  }
  \subfigure[Eight Nodes]{
    \includegraphics[width=3.25in,clip]{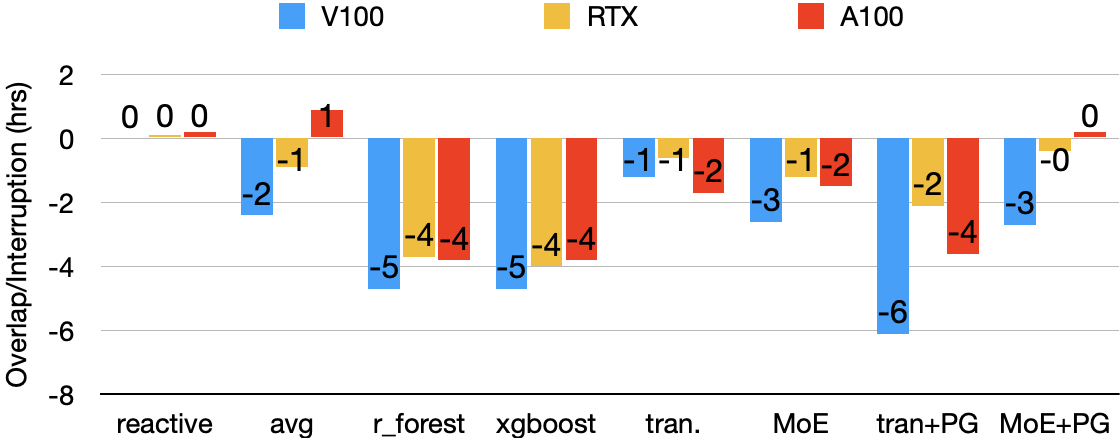}%
    \label{fig:eight-low}
  }
  \caption{Average Overlap of a Pair of 48-hour Eight-node Jobs on the V100, RTX, and A100 GPU Clusters with Light Load.}
  \label{fig:low-load}\up\up
\end{figure*}

Although the ensemble learning and RL-based methods can efficiently reduce the interruption when the machine is heavily or medium loaded. 
They inevitably introduce overlap between jobs when the machine is lightly loaded, where jobs can easily get compute nodes without a significant long wait time.
Figure~\ref{fig:low-load} shows the overlap of all methods with 1-node and 8-node jobs across the three clusters.
Given the 48-hour wall clock time, a few hour overlap is not a big problem for the proactive provisioner, as the subsequent job needs to load data, checkpoints, and libraries. 
Once it is ready, the current job can be released, and the subsequent job will resume from the latest checkpoint, so there is no redundant computation or wasted node hours.
However, we want the proactive provisioner to be intelligent to avoid overlap if possible.
With this standard, we observe that the ensemble methods and the transformer+PG method introduce a $\sim$2X long overlap as the MoE+DQN method.

To balance the interruption reduction and overlap across machine load level, \name{} uses the MoE+DQN as its default model. 
We leave transformer+PG as a option for users as an aggressive provisioner, which will be more effective when the machine load is high.